\documentclass[fleqn,12pt]{wlscirep}
\usepackage[utf8]{inputenc}
\usepackage[T1]{fontenc}
\usepackage{bookmark}

\title{Stability of Imbalanced Triangles in Gene Regulatory Networks of Cancerous and Normal Cells}

\author[1]{Abbas Karimi Rizi}
\author[1]{Mina Zamani}
\author[1]{Amirhossein Shirazi}
\author[1,2,*]{G. Reza Jafari}
\author[2]{J\' anos Kert\'esz}
\affil[1]{Physics Department, Shahid Beheshti University, G.C., 1983969411 Tehran, Iran}
\affil[2]{Department of Network and Data Science, Central European University, H-1051 Budapest, Hungary}

\affil[*]{g\underline{\space}jafari@sbu.ac.ir}


\begin{abstract}
Genes communicate with each other through different regulatory effects, which lead to the emergence of complex structures in cells, and such structures are expected to be different for normal and cancerous cells. To study breast cancer differences, we have investigated the Gene Regulatory Network (GRN) of cells as inferred from RNA-sequencing data. The GRN is a signed weighted network corresponding to the inductive or inhibitory interactions. Here we focus on a particular of motifs in the GRN, the triangles, which are imbalanced if the number of negative interactions are odd. By studying the stability of imbalanced triangles in the GRN, we show that the network of cancerous cells has fewer imbalanced triangles compared to normal. Moreover, in the normal cells, imbalanced triangles are isolated from the main part of the network, while such motifs are part of the network's giant component in cancerous cells. 
Our result demonstrates that due to genes' collective behavior the complex structures are different in cancerous cells from those in normal ones.
\end{abstract}
\begin{document}

\flushbottom
\maketitle
%
%
\thispagestyle{empty}


\section*{Introduction}

Cancers are a large family of diseases that involve abnormal cell growth with the potential to invade or spread to other parts of the body \cite{nci2018}. From the reductionist perspective, cancer is known as a disease of the genes. From this perspective, related studies focus on finding particular genes for each type of cancer and, consequently, diagnosing or curing cancer face formidable challenges. On the other hand, from the complexity theory perspective, collective behaviors emerged from the interactions of systems with many interacting units, are not describable solely by knowing the behavior of the system’s building blocks (genes), and we cannot understand what happens at a higher level of organization by just studying how each element works at a lower scale. In other words, we need a holistic point of view to study the collective behavior of the genes.\cite{Dr_1}
The human body contains more than 10 trillion ($10^{13}$) cells, originating from a single one. Cells differ from each other, depending on which genes are turned on.\cite{Cell} The process by which information from a gene is used to synthesize  functional gene products (often proteins) is called gene expression. Today, there are several projects globally, compiling genomic information related to cancers, and recent advances with sequencing technology reveal the high importance of these projects. Despite all the advances in technology and analysis in genome sequences, it seems that cancer remains indomitable to a large extent. While we know some genes play an essential role in specific cancers, we are often far from controlling, let alone curing them \cite{Dr_2,Torabi}.

Gene expressions are not independent.\cite{mina_3} They communicate with each other through regulatory effects, in a sense that some genes can up-regulate or down-regulate the expression level of other genes. These complex interactions between the genes can lead to collective behavior and result in changing the state of the cell. In this scenario, there is a network of interactions, in which each gene is represented as a node, and its regulatory effect on other genes is considered the links connecting it to other nodes. These links can have zero (no effect), positive (up-regulation), or negative (down-regulation) weight, forming a weighted signed network. Such networks are called \textit{Gene Regulatory Networks} (GRN). \cite{barbagenet,Davidson,walhout2011gene,Tieri,Liesecke,Costanzo,old_ver,Editors_1,Editors_2}.

Since the advent of high-throughput measurement technologies in biology in the late 90s, reconstructing gene regulatory networks' structure has been a central computational problem in systems biology \cite{2018arXiv180104087H}. Despite the efforts, the exact causal relationships between each pair of genes are unknown and thus, we use the simplification of considering the network as undirected. 
Furthermore discussion of gene expression and interactions is highly complex, which is why higher-order interactions are expected. One of the simplest interactions of a higher than two orders is a third-order called balance theory\cite{old_ver11}.We use Balance theory as the simplest model that does not consider interactions independent of each other and regards them as triadic interactions \cite{fritz,antal}.

To assess the pairwise interaction network structure, we use a  maximum-entropy\cite{old_ver10} probability model to explore the properties of the GRN. Such maximum entropy models have been widely used in statistical physics, e.g., for Ising type interacting models \cite{2017AdPhy..66..197N,old_ver12}. Physical systems in thermal equilibrium are described by the Boltzmann distribution, which has the maximum possible entropy given the mean energy of the system \cite{PhysRev.106.620,old_ver9}.

\section*{From Real Data to Gene Interaction Network}
\label{sec:data}

The mRNA data(expression level) of 20532 genes in the case of Breast Cancer (BRCA: Breast invasive carcinoma) has been downloaded from \textit{The Cancer Genome Atlas} (TCGA) project \cite{TCGA,TCGA_data}. The data contain 114 normal and 764 cancerous samples, and the measurement of the expression levels has been done with the technique of RNA sequencing (RNA-Seq). We have used the Reads Per Kilobase transcript per Million reads (RPKM) normalized data. RPKM puts together the ideas of normalizing by sample and by the gene. When we calculate RPKM, we are normalizing for both the library size (the sum of each column) and the gene length. 
In the following we had to reduce the number of genes because it was difficult to handle a 20532$*$20532 matrix computationally. For each gene, we have calculated the variance of its expression level over its samples, and finally we have stored the first 483 genes with the highest variance, which is due to more different activity patterns these genes show\cite{link}. Note that there are so-called housekeeping genes that typically get transcribed continually. These genes are required to maintain basic cellular function and are expressed in all cells of an organism under normal and pathophysiological conditions \cite{Eisenberg2003HumanHG,old_ver2,old_ver3}. Some housekeeping genes are expressed at relatively constant rates in most non-pathological situations.

\label{sec;Inference}

Measuring interactions is difficult within a living cell, but measuring abundances of components (mRNA levels) is considerably easier. Therefore, from the experimental data we wanted to reconstruct the gene-gene interactions computationally based on a model, following the practice that collective behaviors in such systems are described quantitatively by models that capture the observed pairwise correlations but assume no higher-order interactions \cite{2006Natur.440.1007S}. By assuming a maximum entropy pairwise model, we were looking for the interaction matrix $J$, whose every element $J_{ij}$ is the strength of the net interaction between gene $i$ and gene $j$. In other words, the strength and sign of the interaction represent the mutual influence on each other of a pair of genes' expression levels.
From the maximum entropy probability distribution, we have constructed the energy function, which in this case is an Ising-like model with long-range Ferro- as well as antiferromagnetic couplings, which may lead to frustrated triangles. The energy function for our problem can be written as:
\begin{equation}
    H = -\sum_{i<j} J_{ij}S_iS_j,
    \label{hamiltonian}
\end{equation}
where the expression level of gene $i$ as a continuous real-valued variable (a Gaussian field) is represented by $S_i$. Using the energy function above, we can write down the Boltzmann equilibrium distribution as:
\begin{equation}
    P(\{S_i\}) = \frac{1}{Z} \exp{(-\sum_{i<j} J_{ij}S_iS_j)}.
    \label{prob}
\end{equation}
$Z$ is the partition function, and we have subsumed temperature into the couplings $J_{ij}$ without loss of generality. The interaction matrix, $J$, is not known, and we wanted to \textit{learn/ infer} it \cite{2017AdPhy..66..197N} from the experimental data. We want to infer all the $J_{ij}$ as the parameters of our model. To this end, we have restricted ourselves to a probabilistic model with terms up to second order, which we have derived for continuous, real-valued variables. In other words, our model is constrained to generate the first and the second moments which are exactly the same as what we find from the experimental data.\cite{10.1371/journal.pcbi.1004182} Thus, $P$ must maximize the Gibbs-Shannon entropy to infer the parameters of the model.

\begin{equation}
    S[P] = -\sum_{i<j} P(\{S_i\})\ln{P(\{S_i\})}.
    \label{entropy}
\end{equation}
Using Lagrange multipliers, it can be shown \cite{10.1371/journal.pcbi.1004182} that the desired model is a multivariate Gaussian distribution, twice of its covariance is minus the inverse of the interaction matrix.
\begin{equation}
    P(S;\displaystyle \langle S \rangle;C) = \frac{e^{-\frac{1}{2}(S-\langle S\rangle)^TC^{-1}(S-\langle S \rangle  )}}{(2\pi)^\frac{L}{2}\det{(C)}^{\frac{1}{2}}}.
    \label{multi_gauss}
\end{equation}
So, within this approximation, we can write $J_{ij}=-C^{-1}_{ij}$. $L$ is the number of genes based on which we have built the distribution. The elements of the matrix $J$ are, by definition, the effective pairwise gene interactions that reproduce the gene profile covariances \cite{Lee2014,old_ver4} exactly while maximizing the entropy of the system. The inverse of the covariance matrix, $C^{-1}$, which is commonly referred to as the precision matrix, displays information about the partial correlations of variables. In practice, the precision matrix can be estimated by simply inverting the sample covariance matrix, if a sufficiently large number of samples are available. In our study, due to the lack of enough samples, the inverse of the covariance matrix has been obtained by means of the Graphical Lasso (GLasso) algorithm \cite{10.1093/biostatistics/kxm045}. GLasso is an algorithm to estimate the inverse of the covariance matrix from the observations from a multivariate Gaussian distribution. In statistics and machine learning, lasso (least absolute shrinkage \cite{10.1007/978-3-642-38342-7_15} and selection operator) is a regression analysis method that performs both variable selection and regularization in order to enhance the prediction accuracy and interpretability of the statistical model it produces. G-Lesso sparse the network in such a way that it does not disrupt the overall properties of the network. In sparsing a matrix, One of the problems is that the threshold method in the network is severe. In this way, in networks the threshold may eliminate weak links in favor of solid links. But we know that some links are fragile, and their share in the network is very high. For example, it connects part of the network to another part, but it can be a strong link between the network and the node that does not matter to us. The threshold method eliminates the important weak link that connects the two network parts—in contrast, keeping a strong link connected to the trivial part of the network. We know that removing a strong link that is only connected to an insignificant node does not destroy the network properties while removing a weak link that affects the network properties, G-Lesso is wary of such issues.
\newpage

Following are step by step calculations in brief:
\begin{itemize}
  \item Import Row data from TCGA Database,The mRNA data(expression level) of 20532 genes.
  \item Dimension reduction, keep genes with the highest variance (483 genes).
  \item Calculate the covariance matrix of genes (483*483).
  \item Calculate $J$, inverse of the covariance matrix by G-Lasso\cite{old_ver5} approach to make it sparse,
  with penalti = 0.09.
  \item Calculate Energy-Energy matrix. 
\end{itemize}

All of the calculations have been done in Python and MatLab. All codes and results are available upon request\cite{link}.

\section*{Frustration in Interaction Network}
\label{sec;Frustration}
The positive (negative) value of the interactions implies that increasing (or decreasing) a gene’s expression results in up-regulating (down-regulating) of the other gene(s)’s expression(s), respectively. $J$ is the generalized adjacency matrix~\cite{barabasi}, representing the presence and weight of a link. $J_{ij}$ is the strength of the interaction between gene $i$ and gene $j$ or in network terms, the weight of the link $i-j$. 

Let us now consider the local triangles; Groups with three interacting genes forming a triangle of interactions in the network. The triangle $\Delta(i,j,k)$ is defined as balanced if the sign of the product of its links is positive; $J_{ij}J_{jk}J_{ki}>0$, otherwise, the triangle is imbalanced or frustrated; $J_{ij}J_{jk}J_{ki}<0$. We define a triangle to be of type $\Delta_k$ if it contains $k$ negative links. Thus $\Delta_0$ and $\Delta_2$ are balanced, while $\Delta_1$ and $\Delta_3$ are imbalanced \cite{doi:10.1080/00223980.1946.9917275}. The statistics of the analogues of these imbalanced triangles have been shown to be relevant in systems with signed interactions like random magnets \cite{fischer1994} and social networks \cite{2005PhRvE..72c6121A}. 

The notion of balance allows us to define an "energy landscape" for such networks \cite{bilayer,old_ver6}. For a triangle this is:
\begin{equation}
\Delta_{ijk} =  E_{ijk}=-J_{ij}J_{jk}J_{ki}
\end{equation}
and by summing over all the $E_{ijk}$
the energy of the whole network can be obtained \cite{Kula2019}.
\begin{equation}
E_{total} = \frac{1}{N}\sum_{i,j,k=1}^{N} \Delta_{ijk}
\end{equation}
Note that this energy is different from that of (1) and serves to characterize the triangles, while $H$ was used to calculate the interactions from the measured expression strengths.
Energy counts the number of triangles and does not indicate where the triangles are. The correlation between triangles shows which triangle with energy $E_{k}$  has a common link with  which triangle with energy $E_{l}$.

\begin{equation}
C_{kl} = \frac{1}{N}\sum_{i,j,k=1}^{N} \Delta_{kij}\Delta_{lij}
\end{equation}

\begin{figure}[]
    \centering
    \includegraphics[width=1\textwidth]{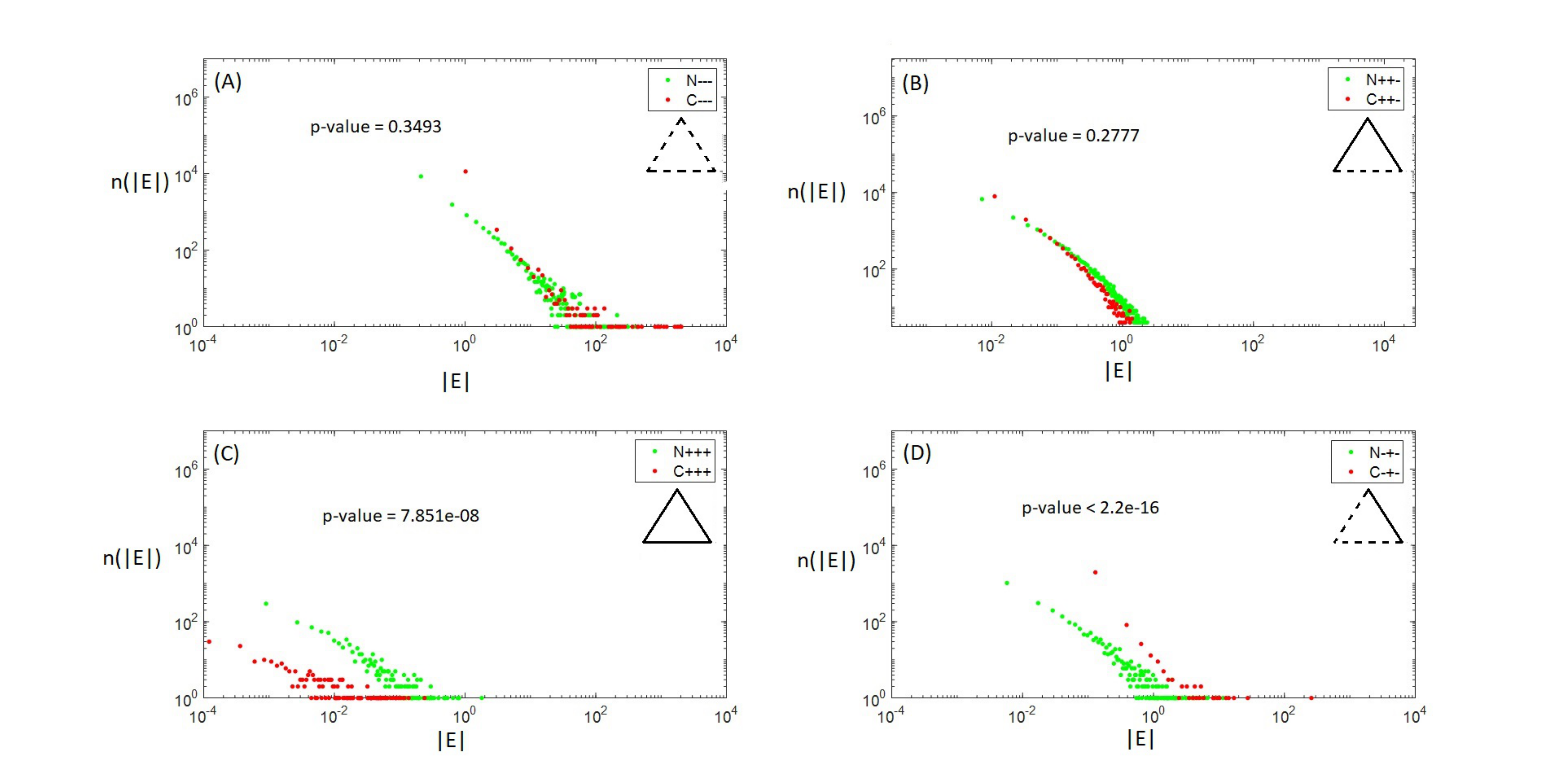}
    \includegraphics[width=1\textwidth]{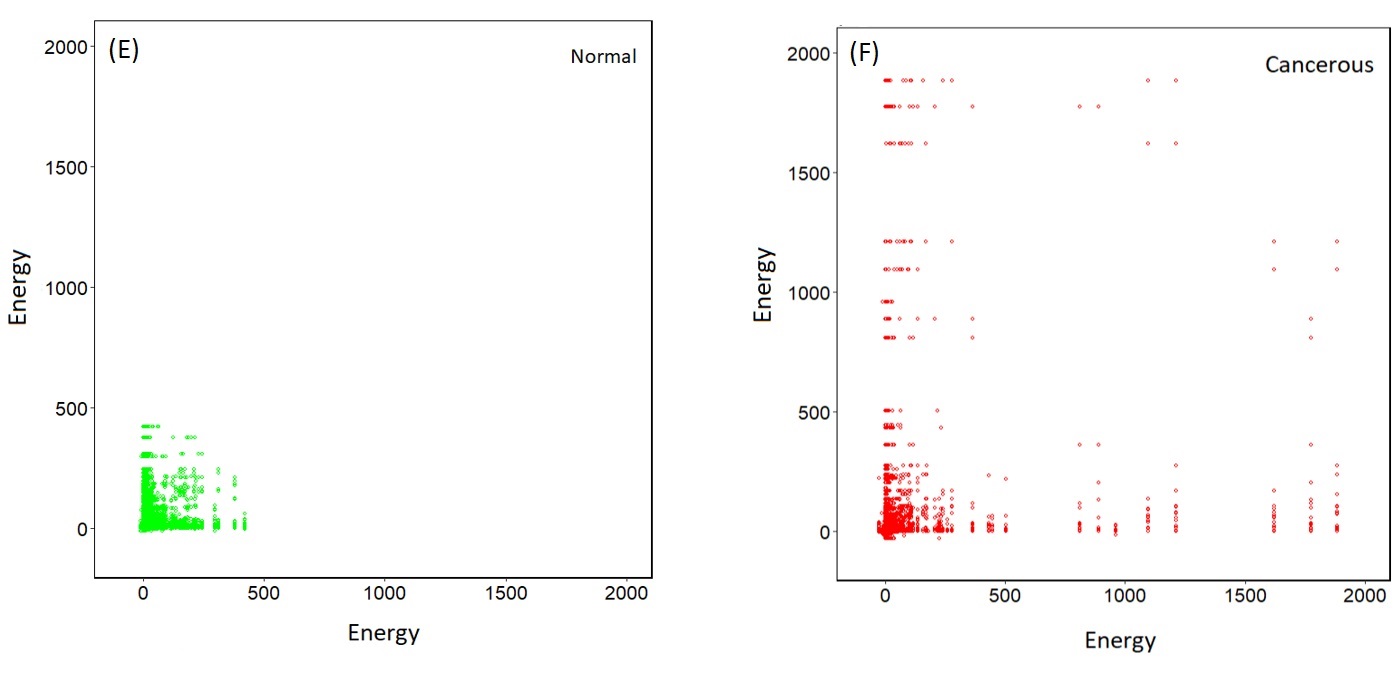}
    \caption{(A-D) Log-log plot of the distributions of triangles vs their absolute energy. All distributions are fat-tailed. Note the differences in the profile of $\Delta_0$ and $\Delta_2$ in cancerous and normal case. (E-F). In cancerous (right) and normal (left) cells triangles with different energies are connected to each other differently. The energy pattern in the normal case is more localized and assortative.}
    \label{fig:fig1}
\end{figure}

This equation answers the question that a triangle with a preferred energy is adjacent to which triangle? result (\autoref{fig:fig1}-E,F) show that in cancer data, high-energy triangles are connected to the rest while in normal data high-energy triangles are not connected to each other. Moreover, in a normal cell frustrated triangles are not part of the main of the network.

\section*{Results}
\label{sec;result}

We have calculated the distributions of the energies of different types of triangles in both cancerous and normal data-sets and observed the following results (\autoref{fig:fig1}). (i) In all the cases, the energy distributions of all types of triangles are fat-tailed. (ii) The distributions of imbalanced (frustrated) triangles, $\Delta_1$ and $\Delta_3$, do not show noticeable differences between cancerous and normal data. 
(iii) In the cancerous network $\Delta_0$ triangles and normal network $\Delta_2$-types are less and the total energy of the cancerous network is lower than that of normal network: 27,239 and  35,984 units, respectively.

In order to see if the effect comes from structural correlations specific to the differences between the normal and cancerous data, we have shuffled the links in the networks. This was carried out by swapping endpoints of randomly selected pairs of links many times, which is a standard procedure to produce degree preserving random reference networks.
The energy difference between the shuffled networks is 280 units which is one order of magnitude less than in the original case. Moreover, the distribution profiles change dramatically for the shuffled network.

\begin{figure}[]
    \centering
    \includegraphics[width=1\textwidth]{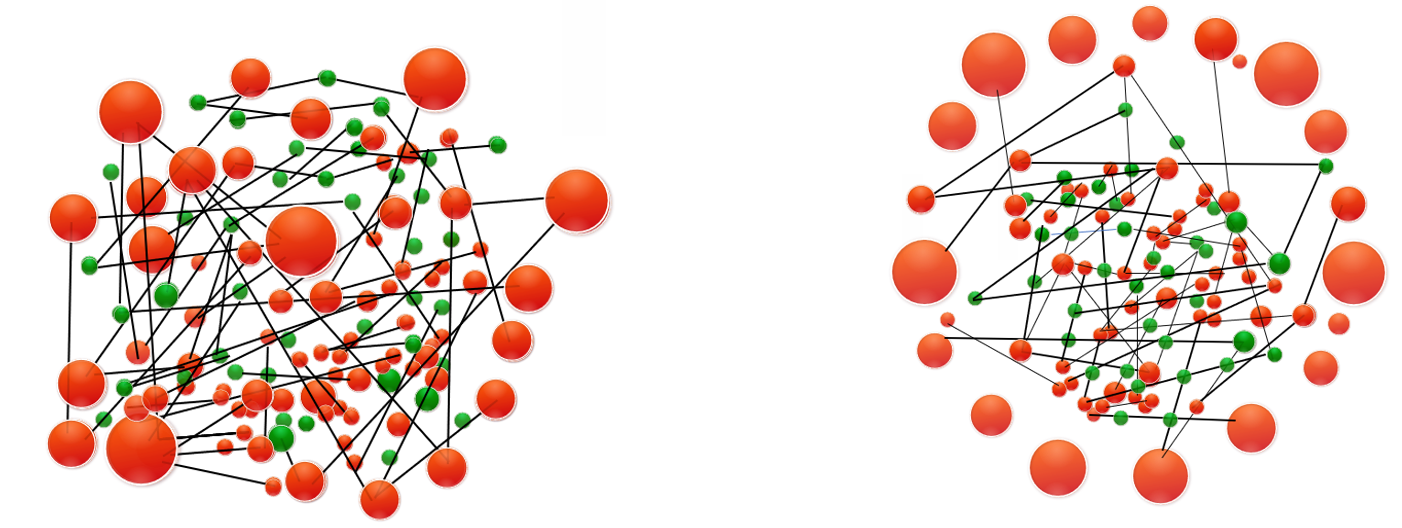}
    \caption{A representation of energy-energy matrix and a schematic diagram of how high-energy frustrated (imbalanced) triangles are distributed in the network of triangles in the normal (right) and cancerous network (left). Compared to the cancerous cell, the normal cell is at a higher energy level, resulting in more likely altering the configuration of the triangles. On the other hand, frustrated triangles (red nods) are more connected to the cancerous triangle network.}
    \label{fig:fig2}
\end{figure}

The next question we have studied was about the distribution of triangles with different energies in the networks and their relationships. For this purpose, we coarse grain the network such that balanced and imbalanced triangles are represented as green and red nodes, respectively. Two coarse-grained nodes are connected if their corresponding triangles have one edge in common. We calculate the energy-energy mixing pattern \cite{2002PhRvL..89t8701N} between the triangles. The plots in \autoref{fig:fig1}-(E-F) shows how many triangles with different energies are connected. Notice that this matrix is rather sparse reflecting that only low number of the triangles have links in common. In the normal network, frustrated triangles are packed together and they form a kind of module while in the cancerous network they have a more heterogeneous pattern of connections and they are mixed with balanced triangles.  Moreover, triangles with higher absolute valued energies are connected to ones with lower absolute valued energies. In both cases, we see triangles with lower energies are more connected to each other. Triangles in the cancerous network do not tend to distribute evenly in a particular region of energy-energy space. Another result is that in both of the networks so many triangles do not have a link in common.

Having more energy for a cell, in this context, means that there is more tendency toward changing the states of the triangles. In the case of cancerous network, we have seen that triangles exhibit a lower chance of being changed. On the other hand, we see frustrated triangles are somehow uniformly distributed in the cancerous coarse-grained network while they are more localized in the normal coarse-gained case. These facts are mimicked in the \autoref{fig:fig2}.
Inspired by the concept of Balance Theory in social science \cite{old_ver7,old_ver14}, we saw that the interaction network of the normal case has more imbalanced (frustrated) triangles and more energy as a consequence. This energy has been defined in a social context giving a good clue to look at the system of genes as a social system. Not only genes cannot live independent of each other, but they also must pay the cost of living together! Note that changing the expression of a gene can have drastic consequences. Our analysis reveals the fact that to get a true picture of biology at the cell level, it is essential to know the connections and their type between the genes.

\section*{Conclusion}
\label{sec;Conclusion}

Cancer has been commonly known as a group of diseases of the genes and there has been a huge effort to find the effective genes responsible for different cancers. Thanks to such reductionist approaches, we now know some specific genes for some cancers. Genes, however, are not independently functioning in the cell and their expressions are strongly correlated with each other. Recently, it has been recognized that the regulatory effects between the genes can be represented by a gene-gene interaction network and the structure of this network is essential in understanding the collective phenomena, which play a role in developing cancer-related studies. Our results contribute to this line of research\cite{old_ver13,old_ver15}. We have presented a formalism, by which we arrived from the data about gene expressions to an interacting network model, where the interactions were inferred using the maximum entropy principle. The resulting signed weighted network \cite{old_ver8} was analyzed from the balanced and imbalanced triangles perspective. We have found significant differences between normal and cancerous cell GRN-s: There are more imbalanced triangles in normal GRN-s than in cancerous ones and the correlations between such triangles are also different in these two networks.Further investigations are indeed valuable to study when the observed differences develop and whether our observations can be used for diagnostic purposes. 

\section*{Acknowledgments}
\label{sec;Acknowledgements}
The authors would like to appreciate Professor Tamás Vicsek and Professor David Saakian for reading the manuscript and their constructive comments.


\section*{Author Contributions}
\label{sec;Author Contributions}
A.K., M.Z. and A.Sh. conceived the model. A.K. and M.Z. did the computations and G.J. supervised the work. A.K., M.Z., A.Sh., J.K. and G.J. analyzed the results. A.K. prepared the manuscript. All authors reviewed the manuscript.

\bibliography{mina.bib}
\end{document}